\documentclass[preprint,proceedings]{rmaa}

\suppressfulladdresses 

\usepackage{paralist}

\usepackage{psfrag,color}

\SetYear{2006}
\SetConfTitle{Massive Stars: Fundamental Parameters and Circunstellar Interactions}

\title{Spectroscopic Analyses of Massive Blue Stars (Galactic or Extragalactic)} 

\author{
  A. Herrero,\altaffilmark{1,2}}

\altaffiltext{1}{Instituto de Astrof\'\i{}sica de Canarias,
  C/ V\'\i{}a L\'actea s/n, E-38200 La Laguna (ahd@iac.es)}

\altaffiltext{2}{Departamento de Astrof\'\i{}sica, Universidad de
La Laguna, Avda. Astrof\'\i{}sico Francisco S\'anchez 2, E-38071.}

\shortauthor{Herrero}
\shorttitle{Spectroscopic Analyses of Massive Blue Stars}

\listofauthors{A. Herrero}
\indexauthor{Herrero, A.}

\abstract{We review recent advances in our understanding of massive stars through the
analysis of their spectra. Improvements in model atmospheres and analysis methods are 
briefly discussed. Results obtained for stars in the Local Group are compared and
the present status of different open questions, like the temperature scale of OB
stars, the Wind Momentum-Luminosity Relation or the stellar rotation, is outlined.}

\resumen{Repasamos los recientes avances en nuestro conocimiento de las estrellas masivas
a trav\'es del an\'alisis de sus espectros. Discutimos brevemente las mejoras en los modelos 
de atm\'osfera y en los m\'etodos de an\'alisis. Se comparan los resultados obtenidos en
estrellas del Grupo Local y se describe el estado actual de diferentes cuestiones que
permanecen abiertas, como la escala de temperaturas de las estrellas OB, la Relaci\'on
entre el Momento del Viento y la Luminosidad o la rotaci\'on estelar.}

\addkeyword{Stars: atmospheres}
\addkeyword{Stars: early types}
\addkeyword{Stars: mass loss}
\addkeyword{Galaxies: Local Group}

\begin{document}
\maketitle

\section{Introduction}
\label{sec:intro}

Massive stars play a central role in our understanding of the Universe. During their
evolution, they inject huge quantities of radiative and mechanical energy that
modify their surroundings to finally explode as supernovae, sometimes generating
Gamma Ray Bursts, and leaving behind neutron stars or black holes. The consequences of
such destiny reach far-lying regions, triggering new star formation epidodes. And the
elemental abundances of very important ions are altered, from He and O to Si and Fe 
through C, N, Na, Mg and Ca, to mention only some of them. They represent for us the 
gateway to the early Universe, as they dominate the spectra of nearby starbursts
and high z galaxies. 

To take adavantage of the presence of massive stars in relation to so many astrophysical
problems we have to be able to explain their structure and
evolution, predict their fluxes at all wavelengths, understand their physics under different
conditions. The way we follow is to try to reproduce 
their observed properties, particularly their spectra, using our model atmospheres. But we have to
do it not only for a limited sample of stars, but for massive stars under all possible conditions,
specially for different metallicities. Fortunately, the Local Group displays a large range 
in metallicity, rendering it an excellent laboratory to study the physics of massive stars.

I'm particulaly happy in this occasion that the first paper in which I could collaborate (a very
small contribution) analyzing stars in the Local Group beyond our Galaxy had Virpi Niemela as one
of the co-authors.  

\section{Model Atmopshere Status}
\label{sec:codes}

Model atmospheres for massive stars have experienced a boost in the last few years,
as a consequence of the work continously developed since the early eighties.
Thanks to this work, we have today different codes available that can be used
for different applications in the field of massive blue stars. Table~\ref{tab:codes}
gives an overview of these codes and their characteristics. Note that these data are
only for a rough overview. Codes may offer several possibilities for a given characteristic,
be coupled together, or the point (for example, the execution time) 
may depend on details of the calculation. Detailed information
can be found in Santolaya-Rey, Puls \& Herrero (1997) and Puls et al. (2005) for
FASTWIND, Hillier \& Miller (1998) for CMFGEN, Pauldrach, Hoffmann \& Lennon (2001) for
WM-basic and Lanz \& Hubeny (2003) for TLUSTY.

\begin{table*}[!t]\centering
  \setlength{\tabnotewidth}{\columnwidth}
  \tablecols{5}
  \setlength{\tabcolsep}{2.8\tabcolsep}
  \caption{A simple comparison of properties of codes suitable for the analyses of
massive blue stars. } \label{tab:codes}
  \begin{tabular}{lcccc}
    \toprule
     & FASTWIND & CMFGEN & WM-basic & TLUSTY \\
    \midrule
    Geometry & Spherical & Spherical & Spherical & Plane-parallel \\
    Wind     &    Yes    &    Yes    &    Yes    &    No          \\
    Blanketing & Approx.  &  Yes    &    Yes    &    Yes         \\
    Rad. Transfer & CMF-Sobolev & CMF& Sob+Cont. &  Observer's frame \\
    T Structure & e$^-$ balance  &  Rad. Eq.    & e$^-$ balance &   Rad. Eq. \\
    Photosphere & Yes    & Approx.   &   Approx.     &    Yes         \\
    Phot. Broadening  & Yes    &    Yes    &    No     &     Yes         \\
    Hydrod.    &  No     &    No     &  Yes (Sobolev) &   No          \\
    Exec. time (h)&  0.3  &   12      &     4     &     4          \\       
    \bottomrule
  \end{tabular}
\end{table*}

Of course, these codes are different and therefore some careful cross-check is
always needed. Results from such comparisons may be found in Puls et al. (2005)
(for a FASTWIND, CMFGEN, WM-basic comparison) with some update in Najarro et al.
(2006) and in Martins, Schaerer \& Hillier (2005a) (for a CMFGEN, TLUSTY comparison).

\begin{figure}[!t]
  \includegraphics[width=1.00\columnwidth]{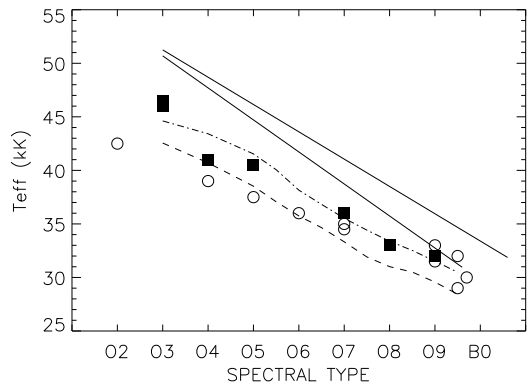}
  \caption{A comparison of temperature scales for O dwarfs and supergiants. 
Solid lines correspond to the Vacca et al. (1996) scales for dwarfs (upper line) and supergiants 
(lower line), while the dash-dotted and dashed lines are the Martins et al. (2005a) scales 
for dwarfs and supergiants, respectively. Solid squares stand for dwarfs from Repolust,
Puls \& Herrero (2004) and open circles stand for supergiants from the same authors. We see
that the Repolust et al. and Martins et al. scales compare well, while the Vacca et al.
scale is much hotter}
  \label{fig:tscaleso}
\end{figure}

\begin{figure}[!t]
  \includegraphics[width=1.00\columnwidth]{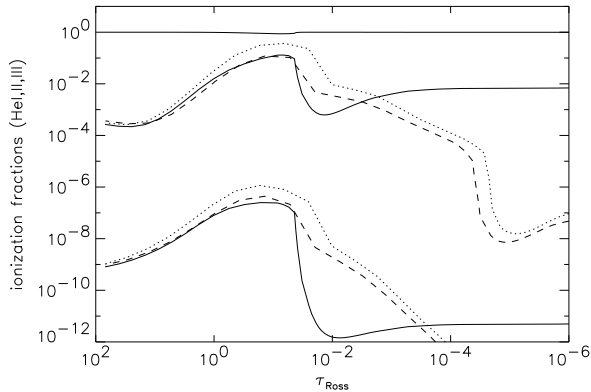}
  \caption{The effect of line blanketing on the He ionization fraction. the calculation is for a model
corresponding to HD 15629, an O5.5 V star. The He ionization fractions increase from HeI to 
HeIII. Solid lines correspond to the blanketed final model and the dashed lines to the
hotter, unblanketed models. Note that both models have very similar ionization fractions,
and thus predict the same spectrum. On the contrary, those of the unblanketed model with 
the same temperature as the blanketed one (dotted line) are different. The reason 
is that the radiation blocked and sent back by the increased opacity raises the temperature of the
blanketed model in the corresponding layers (from Repolust, Puls \& Herrero, 2004)}
  \label{fig:lblank}
\end{figure}

The new models have produced important changes in the O stars calibrations of 
spectral type vs. different physical parameters. Particularly important is the
change in the effective temperature scale of O stars. 
Figure~\ref{fig:tscaleso} shows a comparison of several
temperature scales. We see that the new temperature scales are cooler than
previous ones. The main reason is the introduction of line blanketing, that acts
as an additional opacity source in the UV, as compared to pure H-He model
atmopsheres. Figure~\ref{fig:lblank} (taken from Repolust, Puls \& Herrero, 2004)
explains the effect.

\subsection{The FLAMES Survey of Massive Stars}
\label{sec:flames}

Present model atmospheres relay on a number of assumptions, particularly
that the stellar wind is driven by radiation. As shown by 
Kudritzki et al. (1995)
this implies the so-called Wind Momentum-Luminosity Relation, expressed as
$$ log D_{mom}= x log(L/L_{\sun}) + D_0 $$ where $D_{mom}$ is the product
$(\dot M v_\infty R^{1/2})$ and x and $D_0$ are related to metallicity. To check
this relation and the predictions of the radiatively driven wind theory is one
of the most important tasks presently open in massive stars fields.

To test the WLR at LMC and SMC metallicities is one of the main objectives of the
FLAMES Survey of Massive Stars (P.I., S.J. Smartt). This is an ESO Large-Programme
with 120 hours VLT-FLAMES observing time, plus some extra time to complement the 
observations (mainly with FEROS). Three galactic clusters for comparison and two 
clusters in each of the Magellanic Clouds were observed.
Table~\ref{tab:clusters} quotes some numbers that give an idea of the observational
effort.

\begin{table}[!t]\centering
  \setlength{\tabnotewidth}{\columnwidth}
  \tablecols{5}
  \caption{Observations in the FLAMES Survey of Massive Stars} \label{tab:clusters}
  \begin{tabular}{lcccc}
    \toprule
             & galaxy & WR    &   O &     B \\
             &        & stars & stars & stars \\
    \midrule
    NGC 3293 & MW  & - & -  &  99     \\
    NGC 4755 & MW  & - & -  &  98     \\
    NGC 6611 & MW  & - & 13 &  40     \\
    NGC 2004 & LMC & 1 &  4 & 107     \\
    LH9/10   & LMC & - & 44 &  76     \\
    NGC 330  & SMC & - &  6 & 109     \\
    NGC 346  & SMC & - & 19 &  86     \\
    \midrule
    Total    &     & 1 & 86 & 615 \\       
    \bottomrule
  \end{tabular}
\end{table}

To analyse such a large number of stars requires a lot of time and to lower it,
Mokiem et al. (2005) have presented an automatic fit method for O stars based on the use
of a genetic algorithm that takes advantage of the fast computational properties of
FASTWIND. The algorithm has been applied to the analysis of LMC and SMC O stars
observed in the FLAMES Survey, resulting in the largest number of O stars
homogenously analyzed. As the distance to the stars is known, it is possible
to derive the stellar radius and obtain the luminosities and wind momenta 
for this large number of stars. 

The results for the WLR have been presented in Mokiem et al. (2007b) and indicate a partial
agreement with the theoretical predictions by Vink et al. (2001). Mokiem et al. confirm that
the WLR decreases with metallicity in the expected way: it is highest for the Milky Way, 
intermediate for the LMC and lowest for the SMC, in amounts that also agree with the 
predictions. Moreover, Mokiem et al. confirm that the metallicity dependence of the mass-loss
rate also agrees with theoretical values. Thus, for $\alpha$ in the relation $\dot M \propto Z^\alpha$
Mokiem et al. obtain $\alpha$= 0.83$\pm$0.16, compared with the value by Vink et al.,
$\alpha$= 0.69$\pm$0.10.   

\subsection{Model Atmosphere Status: A word of caution}
\label{sec:caution}

However Mokiem et al. (2007b) could not confirm all the predictions of theory.
While the standard radiatively driven wind theory is well stablished for luminous stars,
some aspects are still controversial. We emphasize here two of them: the possible
presence of wind clumping and the possible breakdown of the theory at relatively
low luminosities, of the order of log(L/L$_\sun \rm ) \approx$ 5.2.

Mokiem et al. found a systematic displacement in the ordinate between the observed and 
predicted WLR for the Milky Way, the LMC and the SMC (a difference in D$_0$, see above). 
The difference corresponds to an overstimation of the observed mass-loss rates by a factor
of about 2. Other authors have also found differences between the observed and predicted WLR, like
Herrero, Puls \& Najarro (2002), Repolust, Puls \& Herrero (2004) or Massey et al.
(2005). The easiest way to explain this would be through wind clumping. As the emission
in H$_\alpha$ (the main mass-loss diagnostic line used by Mokiem et al.) is due to 
e$^-$-H$^+$ encounters, it will be roughly proportional to $\rho^2$, and because
$<\rho^2> \neq <\rho>^2$ assuming a smooth wind in the presence of clumping will 
end in too large values for the wind density and therefore the mass-loss rate
through the equation of continuity. A recent analysis by Puls et al. (2006a), including
a coarse radial dependence, has shown that clumping is present at least in the
denser winds. Because the results from Puls et al. are based on diagnostics that depend on 
$\rho^2$ (H$_\alpha$, IR and radio) it is not possible to derive the absolute value of the
clumping, but only the relative clumpiness between regions where the different
diagnostics form. For the thinner winds, no significant difference could be found.
For the denser winds, the average correction in the mass-loss rates was a factor of
2 (toward values lower than for smooth winds).

The wind diagnostic lines in the UV are usually not affected by wind clumping, because 
the mass-loss rates are derived from absortion resonance lines, being therefore proportional
to $\rho$ and not to $\rho^2$. However, these lines are usually saturated in O stars,
and can therefore not be used for mass-loss determination. Using the P V doublet
at $\lambda\lambda$1118, 1128 Fullerton et al. (2006) find very large factors (of the order of 10 and
more) between the H$_\alpha$ and UV mass-loss rates. The same factors are found by other
authors (see references in Fullerton et al.) and in B supergiants (Prinja et al., 2005).
However, while clumping may also play a role here, and even offer an explanation if we
adopt extreme clumping factors (see also Puls et al., 2006b), 
such huge differences probably need an extra effect 
to explain the low ionization fraction found for P IV and P V, like
the inclusion of X-rays ionization or problems with model atoms.

The second problem to which we refer, the possible breakdown of the radiatively
driven wind theory, has been noticed by Bouret et al. (2003) in NGC 346 stars and 
Martins et al. (2004) in N81, who showed that for SMC stars with relatively low
luminosity the derived mass-loss rates were much lower than predicted. Later, Martins
et al. (2005b) have shown that the effect is present also in Galactic stars, and therefore
metallicity cannot be claimed to be reason for the breakdown. Martins et al. attribute
it to a decreased line force parameter $\alpha$ (not to be confused with the
$\alpha$ exponent in the $\dot M$-Z relation above), but the reason for it is unknown.

\section{Massive blue stars in the SMC and LMC}
\label{sec:near}

The study of stars in the Magellanic Clouds provide us with very important relations, namely
those of the stellar parameters with metallicity. Of course, the FLAMES
Survey of Section~\ref{sec:flames} represents also an important contribution here.
The first relation we may check for variation with metallicity is the T$_{\rm eff}$ vs 
spectral type relation. Two major works have been published recently, that of
Massey et al. (2004, 2005) and that of Mokiem et al. (2006, 2007a). Both works show a
qualitative agreement: stars in the SMC are hotter than their counterparts in the
Milky Way, and stars in the LMC tend to lie in between (although not in a very clear way),
results being similar for dwarfs and supergiants. However, we should note that the
intrinsic scatter is very large at any spectral subtype. For example, at O7, the
SMC stars by Mokiem et al. (2006) enclose in temperature their LMC and Galactic counterparts.

Note that the agreement between Massey et al. and Mokiem et al.
is to be expected: they use the same techniques and the same code (FASTWIND) for
their analyses. On the contrary, while for Massey et al. (2005) the SMC O4-O5 dwarfs are 
hotter than their Galactic analoges, for Bouret et al. (2003) the opposite is true.
This is interesting, as Bouret et al. use CMFGEN to analyse the stars. However, they
only analyzed six stars in NGC 346, and thus we shouldn't give this difference a strong
statistical value. But it is clear that further work is needed before we consider the
temperature scale of O stars well stablished at different metallicities.

\begin{figure}[!t]
  \includegraphics[width=1.00\columnwidth]{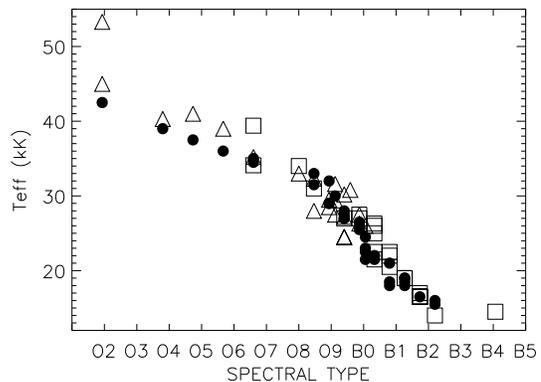}
  \caption{The temperature scale for O and B supergiants in the Milky Way (solid circles, Repolust
et al., 2004), the LMC (triangles, Evans et al., 2004, Mokiem et al., 2007b) and the 
SMC (squares, Trundle \& Lennon, 2005, Mokiem et al., 2006).
The O supergiants are hotter at lower metallicities,
and the trend seems to be present until spectral type B1. For later stars, there is no 
clear trend anymore. Compare with Figure~\ref{fig:tscaleso}}
  \label{fig:tscalesob}
\end{figure}

For the B supergiants, the situation is slightly different. The scales derived by Crowther et al.
(2006) for Galactic stars, Evans et al. (2004) for LMC and Trundle \& Lennon (2005) for 
SMC ones agree, and there is no clear systematic difference between stars from different
metallicities at a given spectral type, although the trend towards hotter types for lower
metallicities seems to be present until spectral type B1. 
Figure~\ref{fig:tscalesob} gives an overview of the
temperature scales of Galactic and Magellanic clouds O and B supergiants, with data from
Repolust, Puls \& Herrero (2004) and Mokiem et al. (2006, 2007b) for O supergiants, and
from Crowther et al. (2006), Evans et al. (2004) and Trundle \& Lennon (2005) for B supergiants.

Another important parameter of massive stars that has to be studied with metallicity is
rotation. Rotation may strongly influence the evolution of massive stars (Maeder \& Meynet,
2000) and a metallicity dependence is predicted for their rotational velocity: because
stellar winds are larger for higher metallicities, therefore producing a larger loss of
angular momentum at the surface as compared to stars of lower metallicity, 
the fraction of fast rotating stars should decrease with increasing metallicity.
 
This is supported by the finding by Maeder et al. (1999) who studied stellar clusters in
the Galaxy, the LMC and the SMC and found a clear trend towards a larger
fraction of Be stars for clusters of lower metallicity. Recent studies by Heap et al. (2006)
and Mokiem
et al. (2006) confirm a large N and He enhancenment in the stars of the SMC, in agreement with theroetical
expectations of CNO contamination induced by rotational mixing. However, the same study by
Mokiem et al. does not find clear evidence of a larger initial rotational velocity in the
stars of the SMC, as compared to the Galactic stars.

An additional point that has to be considered is the possible existence of extra line broadening
that is attributed to rotation while being due to other effects, like large or meso-scale 
turbulent motions. Ryans et al. (2002) have shown that a large fraction of the line broadening
in B supergiants is due to turbulent-like motions, while Sim\'on-D\'\i{}az \& Herrero (2007) have recently 
shown, using the Fourier transform method that line broadening  in O stars has also a turbulent-like
contribution. While this extra broadening is comparable in dwarfs and supergiants, it is
more important in the latter due to the comparatively smaller rotational and collisional broadening
mechanisms.

Finally we will refer here to advances in the mass discrepancy problem. In the Galaxy, the recent
works by Herrero et al. (2002), Repolust et al. (2004) and Mokiem et al. (2005) show that
there is no evidence for a systematic mass discrepancy as that originally described by Herrero et al. (1992), 
although some small
problems may still be present for stars with M$_{\rm spec} \approx$ 20 M$_\sun$ or lower, or for some
stars that seem to occupy a line parallel to the 1:1 relation in the M$_{\rm spec}$ vs M$_{\rm evol}$ 
diagram, but a slightly too large M$_{\rm evol}$ values. However, the evidence for these
two aspects is limited and they should not be considered as firmly stablished.

The situation is different and more complicated when the analyses of stars in the Magellanic Clouds
are considered. In the LMC, the results from Massey et al. (2004) and Mokiem et al. (2007a) seem 
to indicate the same difficulties that we have just pointed out for the Galactic stars. In addition
Massey et al. find a strong mass discrepancy for very hot stars, as do
Mokiem et al. for 2 of their 4 O2 objects, with the other two reaching the 1:1 line
only because of very large error bars. Therefore, Massey et al. (2004) and Mokiem et al. (2007a)
show some reminiscences of the mass discrepancy in the LMC, but at least their results are
consistent. However, while Mokiem et al. (2006) do not find evidence of mass discrepancy in the
SMC (even the limited problems that we have refereed above are not present or appear in milder form),
Heap et al. (2006) find a large mass discrepancy in their SMC stars. More work will be needed for a 
final word, but we note here again that Mokiem et al. and Massey et al. use the same code
(FASTWIND), while Heap et al. have used TLUSTY.

Herrero \& Lennon (2004) suggested that the abundance and mass patterns found in massive blue
stars could be explained by a combination of rotation and mass-loss. According to these authors,
the evolution of very massive stars (initial masses of the order of 50-60 M$_\sun$ or more)
is dominated by mass-loss, while for the lower end of the massive stars (masses in
the ZAMS of 20-25 M$_\sun$ or less) peculiarities in the evolution (like particular abundance
patterns) would be due to rotation. For stars of masses in between, the particular
characteristics of each star would determine the subsequent evolution. If this scenario is
correct, rotation could play an increasingly important role compared to mass-loss as metallicity decreases.

\section{Massive blue stars in nearby spiral galaxies}
\label{sec:spiral}

The analysis of massive blue stars in spiral galaxies offers the opportunity to study the 
evolution of these stars under different conditions, as in the Magellanic Clouds, but in addition
we can also study the structure and abundance patterns of these galaxies similar to our own.

The two major works in this context are those by Urbaneja et al. (2005a,b) in NGC 300 and
M33. In M33 Urbaneja et al. (2005b) use observations of massive blue stars obtained with
the WHT and the Keck-II telecopes, at R$\approx$2500-5000 and SNR $\geq$ 100. Analyzing a total
of 11 stars, Urbaneja et al. could determine the radial O/H abundance gradient, that
agrees well with that found in HII regions by Vilchez et al. (1988), although it cannot 
solve the question whether the gradient decreases linearly with constant slope or has two zones, 
with a strong fall of the O abundance in the innermost part and a small, nearly horizontal slope in
the outer regions. The stellar Mg and Si abundances follow the trend of the O abundances
but with shallower gradients, a difference that, if confirmed, cannot be easily explained 
by present models. Taken together, the $\alpha$-elements abundances indicate solar metallicities
for the central regions of M33. Finally, an interesting comparison with type I Planetary
Nebulae indicates that there has been no significant O contamination over the past 3 Gyr. 

In NGC 300 Urbaneja et al. (2005a) analyzed observations taken with the VLT-FORS2 (Bresolin et al.,
2002a,b). These observations are of lower SNR and resolution than those
in M33 becasue of the larger distance to NGC 300 (about 2.2 Mpc), and forced
changes in the analysis technique. An important lesson is that we can derive 
the stellar parameters and abundances of the brightest B supergiants in other galaxies
even with resolutions of R$\approx$1000, thanks to the relatively few lines present in
the spectra of these stars (although the good SNR is still needed, even more than
at higher resolution or metallicity). The large distance of NGC 300
has also made impossible to obtain directly {\it Te} for the HII regions in NGC 300, so that
the O abundances in these regions were derived using different calibrations. The comparison
with the stellar data favours the calibration by Denicol\'o, Terlevich \& Terlevich (2002), 
that gives a similar value for
the radial O/H gradient and the central O/H abundance to those of Urbaneja et al.
(-0.033$\pm$0.026 dex kpc$^{-1}$ and 8.57$\pm$0.13 dex). 
In this case the Si and Mg gradients agree with the O one.

Taken together we see that both works give consistent
results: the O abundances derived from B-supergiants trace a gradient that is consistent
in both galaxies with studies from HII regions and give central abundances lower than
previously accepted.

\begin{figure}[!t]
  \includegraphics[width=1.00\columnwidth]{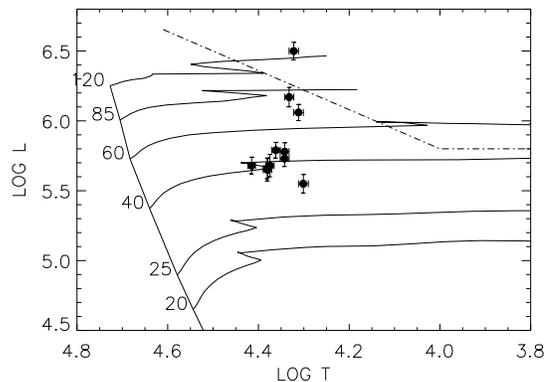}
  \caption{The HRD of the M33 stars observed by Urbaneja et al. (2005b). Tracks are non-rotating
tracks by Schaller et al. (1992) for Galactic metallicity. The dashed-dotted line represents the
Humphreys-Davidson limit. Star 110A is the point above that limit, and B133 is the point
immediately below it. While the luminosity of B133 should be corrected by the presence of
a close companion (and its UV spectrum shows a peculiar SiIV profile), 110A seems to be
an isolated, very luminous star}
  \label{fig:hrd}
\end{figure}

However, the study of stars in external galaxies is subject to numerous problems.
One of the most important is stellar crowding, and is illustrated in 
Figure~\ref{fig:hrd}, where we show a HRD for the B supergiants analyzed by 
Urbaneja et al. (2005b). Two of the stars (110A and B133) are very close to the 
Humphreys-Davidson limit, with one of them (110A) clearly above it. 
In images taken with the HST-STIS accquisition camera we can see that B133 has a
close companion star. Two similar stars are within the circle of 1 arcsec diameter
that represents the seeing disk under which the stars were observed from the
ground. Therefore, the luminosity derived for B133 is too large, and the star, still
very luminous, has a luminosity that does not challenge the theory of stellar structure.
Moreover, an indication of this companion could be seen in the UV spectrum of B133, that was
peculiar (Urbaneja et al., 2002). But this is not the case for 110A, for which no 
peculiarity was found in the UV spectrum, nor in the optical (except for emission in
the Balmer lines), nor in the HST accquisition image. Therefore, 110A may be one of
the most luminous known stars, and even an example of a higly structured atmosphere
leading to {\it porosity} as suggested by Owocki (2004), 
although the possibility of an even closer companion
(or companions) than in the case of B133 has always to be considered.

\begin{figure}[!t]
  \includegraphics[width=1.00\columnwidth]{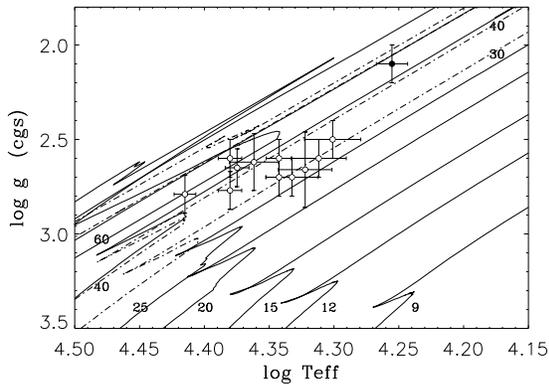}
  \caption{T$_{\rm eff}$- logg diagram of the stars in M33 analyzed by Urbaneja et al. (2005b),
together with the position of UIT 005 (the solid symbol). Temperatures and gravities have 
been derived from the analysis of the spectra. Labels correspond to the initial
stellar masses. Solid lines are for Meynet \& Maeder (2003) tracks for solar metallicity, and
dash-dotted lines are for Meynet \& Maeder (2004) tracks for Z= 0.008. The advanced position
of UIT 005 is clear in the figure}
  \label{fig:uit}
\end{figure}

To find some particularly interesting objects is one of the appealing possibilities
of the extragalactic stellar spectroscopy. For example, we have found that UIT 005,
a B2.5 supergiant in M33 is a very evolved object, with a very low O/H content
indicating a very evolved evolutionary status. The spectroscopic analysis gives
an actual mass of 41$\pm$10 M$_\sun$, while the evolutionary mass is in the
range 33-44 M$_\sun$. Figure~\ref{fig:uit} gives an idea of the comparison
of UIT 005 with evolutionary tracks and with other stars in M33. Clearly, the
position is consistent with a spectrum showing weak O lines as a consequence of advanced
evolution.

The full power of the extragalactic stellar spectroscopy will only be reached when 
we are able to perform extense analyses of stars all over the surface of the
host galaxies. To this end, we have started a program to observe stars over
M33, to trace a 2D map of stellar abundances in this galaxy. This way, we expect
to be able to assess the detailed abundance pattern in M33, indicating possible 
differences between arms and interarms regions, or between the
center and the outskirts of the galaxy. This project is presently being prepared
to be accomplished with the multiobject spectrograph OSIRIS to be attached at
the GTC. We hope to start this program early 2008, as soon as the GTC begins regular
operations.

\section{Acknowledgements}
I would like to thank the organizers of the workshop for the invitation to give this review, 
and Joachim Puls and Miguel Urbaneja for very useful comments. This work has been
funded by the Plan Nacional de Astronom\'\i{}a y Astrof\'\i{}sica of the spanish
Ministerio de Educaci\'on y Ciencia through project AYA 2004-08271-C02-01.

\end{document}